\documentclass[preprint,12pt]{article}
\usepackage{amssymb}\usepackage{amsmath}
\usepackage{epsfig}
\usepackage{subfigure}

\begin{document}
\def\RR{X_{rr}}
\def\IR{X_{ir}}
  \def\II{X_{ii}}
  \def\SR{X_{sr}}
 \def\SI{X_{si}}
 \def\SII{(X_{si})^2}
 \def\SS{X_{ss}}
 \def\S{X_{s}}
  \def\I{X_{i}}
   \def\R{X_{r}}
   \def\EL{\scriptscriptstyle L}
   \def\EG{\scriptscriptstyle G}
\title{SIR model with local and global infective contacts: A deterministic approach and applications}

\author{Alberto Maltz$^{1 +}$  and Gabriel Fabricius$^{2 *}$ \\
\\
\small{$^1$ Departamento de Matem\'atica} \\
\small{Facultad de Ciencias Exactas, Universidad Nacional de La Plata,}\\
\small{CC 72, Correo Central, 1900 La Plata, Argentina}\\
\\
\small{$^2$ Instituto de Investigaciones Fisicoqu\'{\i}micas Te\'oricas
         y Aplicadas,} \\
\small{ Facultad de Ciencias Exactas,
         Universidad Nacional de La Plata,}\\
\small{ cc 16, Suc. 4,
         1900 La Plata, Argentina} }

\maketitle

\begin{abstract}
   An epidemic model with births and deaths is considered on 
a two dimensional $L\times L$
    lattice. Each individual can have global infective contacts
    according to the standard SIR model rules or local 
    infective contacts with its nearest neighbors.
    We propose a deterministic approach to  this model and 
verified that there is 
a good agreement with the stochastic
    simulations 
for different situations of the disease transmission 
and parameters corresponding to pertussis and rubella in the prevaccine era.
\end{abstract}


\addvspace{15pt}
\small{$^+$ {\it e-mail:} alberto@mate.unlp.edu.ar }

\small{$^*$ {\it e-mail:} fabricius@fisica.unlp.edu.ar }

\section{Introduction}
\label{intro}

  Mathematical modeling of infectious diseases has become an area
  of increasing interest in the last decades 
  \cite{libroAM, libroKR,reviewscience2015}.
The development and use of mathematical models are a
  powerful tool to understand the complex problem of infectious disease
transmission. After the success of the simple SIR compartmental
model in the description of the basic and common features of the
transmission process \cite{kermack, libroAM}, the models have become
more complex and specific for the different infectious diseases in order
to help in the evaluation and design of control strategies
\cite{Hethcote9799, modelVIH, nos-adol-boost}.
These complexities may include, among others, age structure of the population,
immune status of the individuals, structure of the social contacts,
spatial heterogeneity, etc.

Wherever possible, deterministic compartmental models are usually
chosen because they are simple to solve numerically, the interpretation
of results is direct, and the level of complexity may be increased
gradually by adding new compartments.
However, from the beginning of infectious disease modeling,
the importance of stochastic effects has been mentioned \cite{bartlett56}.
In more recent work, the significance that a stochastic treatment
could have in the transmission of some diseases such as pertussis
has been highlighted \cite{science-rohani-99}.
In particular, the randomness and heterogeneity of contacts are one of
the intrinsically stochastic aspects of the contagion process
and knowing the accuracy that
can be obtained with a deterministic approach to the problem
is not a trivial matter.

In ref.\cite{lebo} a SIRS spatial stochastic model
 having only local infective contacts is considered on a lattice
and a PA scheme to approach the dynamic is developed 
(some details are described in section \ref{sectionDA}).
In ref.\cite{verdasca} another model on a lattice is proposed, where
local and global contacts are considered with the aim of emulating
the more realistic situation of social relations where some ``fixed''
persons are frequently contacted (the neighbors on the lattice)
and some other unknown people are met by chance (the global contacts).
In their model, a parameter $p$ characterize the weight of global contacts
and, in this way, the influence of the degree of locality or globality
of the contacts on the dynamics of the disease transmission
is analyzed by changing a single parameter.
 Other authors have characterized the stochastic fluctuations of
this model as a function of $p$ for a wide range of model parameters
that include several infectious diseases \cite{simoes}.
Dottori {\it et al.} \cite{dottori} studied the quasi-stationary state (QSS)
of the model for parameters corresponding to pertussis disease
and quantified the relation between the effective transmission of
the disease and the correlation of susceptible individuals with their infected
neighbors.

In the present work we propose a deterministic approximation (DA) 
to the stochastic model with local and global contacts (SM) mentioned above,
where the local contacts on the square lattice are treated here under 
  a PA scheme based on the one used by Joo {\it et al.} \cite{lebo}. 
We compare the
predictions of our proposed DA with the results of simulations performed 
with the stochastic model for parameters corresponding to two
infectious diseases: pertussis and rubella. We make this comparison 
for different situations of the disease transmission dynamics: 
the epidemic spread, the endemic
state of the disease and 
dynamic perturbations that may arise from control measures implemented against
the disease. In all the cases studied, the agreement between the 
results obtained 
from the SM simulations and DA was  very good.

\section{Stochastic model}
\label{sectionSM}
We consider an $L \times L$ square lattice  under periodic conditions 
where we identify sites with individuals. Each one of the $N=L^2$ individuals may 
be in one of the three epidemiological states: $S$, $I$ or $R$ (susceptible, infected
or recovered). The state of an individual may change through the following processes:
infection ($S\rightarrow I$), recovering ($I\rightarrow R$), 
death and birth ($S\rightarrow S$, $I\rightarrow S$, $R\rightarrow S$). 
Infections occur through infective contacts among susceptible and infected
individuals. We define an infective contact (IC) as a contact between two
individuals such that if one individual is susceptible and the other
infected, the former becomes infected. We assume that
an individual at a given site 
has an IC with a randomly chosen individual on the lattice 
with probability rate  $p \beta$, and with one of their four nearest
neighbors with probability rate $(1-p) \beta$.
Local contacts represent the contacts in the circle of stable
relations of an individual, while global random contacts represent
people met by chance (for example, on a bus, at the supermarket, etc.).
By changing $p$, we may change the degree of ``locality" of the contacts
in the system. The case $p$=1 corresponds to the classical SIR model
(uniform mixing) where an individual may have an IC with any other
individual in the system with the same probability rate $\beta$.
Recovery from infection in this model is the same for every site
and occurs at a probability rate $\gamma$.
Deaths are assumed to be independent of the individual's state
and occur at the same probability rate $\mu$.
When an individual dies at a site, another individual
is born simultaneously at this site in order to avoid
empty sites during the simulation.
The processes defined above are Markovian, so, the defined
probability transition rates between states for each site at a given
time control 
the dynamical evolution of the system.
Stochastic simulations are performed using Gillespie
algorithm \cite{gillespie}. 
The algorithm gives a sequence of times and
the corresponding states 
where two consecutive states
differ by a single process that occurs at a given site.
The process and the time when it takes
place are generated from simple rules and two random numbers
(see Ref. \cite{gillespie} for details).

For a detailed description
of the model and the implementation of the simulation
algorithm, see ref.\cite{dottori}.

\section{A deterministic approach}
\label{sectionDA}

       We consider an $L \times L$ lattice
       as in section \ref{sectionSM}. At each time $t$, the individuals must be in one
        of the three states: $S$, $I$ or $R$ (susceptible, infected or recovered).
        We define $C_s$, $C_i$ and $C_r$ as the total number of individuals in those states.
         We denote $U_a=H$ when the state of the individual $a$ is $H$ 
  (for example, $U_a=S$).
        Each pair $ab$ of (horizontal or vertical) neighboring individuals
        can be in six possible states : $SS$, $SI$, $SR$, $II$, $IR$, $RR$.
We remark that $SI$ involves the cases $(U_a=S,U_b=I)$ and
        $(U_a=I,U_b=S)$;
       and the same holds for $SR$, $IR$.
       Let $C_{ss}, C_{si}, C_{sr}, C_{ii}, C_{ir}, C_{rr}$ be the total number
      of each type pair of neighboring individuals.
    Our unknowns, arranged in the vector {\bf X}=($\S,\I,\R,\SS,\SI,\SR,\II,\IR,\RR$),
    are the preceding quantities normalized by $N$.
    They are functions of the time $t$ and are linked by

   \begin{equation}
    \S + \I + \R = 1  \ \     \label{L1}
    \end{equation}
 \begin{equation}
    \SS + \SI + \SR + \II + \IR + \RR = 2 \  \     \label{L2}
   \end{equation}
\begin{equation}
 \S = \frac{ 2\SS + \SI + \SR}{4}  \ \       \label{L3}
 \end{equation}
\begin{equation}
 \I = \frac{2\II + \SI + \IR}{4} \  \        \label{L4}
\end{equation}

     ~(\ref{L1}) and ~(\ref{L2}) hold because there are  $N$ individuals and $2N$ pairs of neighboring individuals.
    To obtain ~(\ref{L3}) (and analogously~(\ref{L4})) observe that
    each
    individual in susceptible state $S$ belongs to four pairs of neighboring individuals of type
   $SS$, $SR$ or $SI$
   but in their total contribution $4C_{s}$, the $SS$ pairs are counted
   twice.


    In the DA equations that we  propose, the terms related to
   the birth-death and recovery rates do not depend on the
   local structure and are treated as in the classic SIR scheme.
       But in the case of the 
transfer rates among classes due to infections, $r$, 
we make a decomposition
        $r=pr_{\EG}+qr_{\EL}$  being $q=1-p$ and $r_{\EG}$,
      $r_{\EL}$ two different rates that must be defined for
       global an
      local infective contacts respectively.
      The decomposition is motivated by the
   following argument in SM.

    Consider an individual (or a pair, according to the case) in any of the states 
$S$, $SS$, $SI$ or $SR$ at time $t$.
     It can reach, through an infective contact, the new state
       $I$, $SI$, $II$ or $IR$ respectively.
       Denote $A_{\Delta t}$ the event
     ``the individual (or the pair) changes the first state by the second one due to an
    infective contact in the time interval
    $(t,t+\Delta t]$". Then
       $P(A_{\Delta t})= r\Delta t + o(\Delta t)$.
      But $A_{\Delta t}$ is the disjoint union
  of $A_{\Delta t}^G$ and $A_{\Delta t}^L$, the events corresponding to one
global and one local infective contact in $(t,t+\Delta t]$
  respectively.
   Assuming that $r_G$ and $r_L$ are known, we have (according to the SM rules)
   $P(A_{\Delta t}^G)=r_G\Delta t + o(\Delta t)$ and $P(A_{\Delta t}^L)=r_L\Delta t + o(\Delta t)$.
   Finally, by the total probability law,
  $P(A_{\Delta t})=P(A_{\Delta t} |A_{\Delta t}^G)P(A_{\Delta t}^G) + P(A_{\Delta t}|A_{\Delta t}^L)P(A_{\Delta t}^L) = (pr_G + qr_L)\Delta t + o(\Delta t)$.

     Let $ab$ and  $bc$ be two different pairs of neighboring individuals sharing $b$.
      The key fact of the PA scheme in ref.\cite{lebo} (with only local infective contacts)
       consists in the approximation of the  probability that the triplet $abc$ reaches a given state
   $(U_{a}=H_1,U_{b}=H_2,U_{c}=H_3)$ at time $t$ by
   \begin{equation}
  P_{t}(U_{a}=H_1,U_{b}=H_2,U_{c}=H_3)\simeq \frac{P_{t}(U_{a}=H_1,U_{b}=H_2)P_{t}(U_{b}=H_2,U_{c}=H_3)}{P_{t}(U_{b}=H_2)}  \label{lebo}
  \end{equation}

     In order to apply the above     approximation to the treatment of the
 local contacts in DA we consider
      the natural correspondence between fractions (our unknowns in DA)
     and probabilities.
    For example $\S$ (the fraction of susceptible individuals at time $t$ in DA)
    can be regarded as the
    probability that a given individual $a$ is of type $S$ at time $t$ in SM,
    and  $\SI /2$ (the fraction of $SI$ pairs at time $t$
     in DA) as
    the probability that a given pair $ab$ of neighboring individuals 
is of type $SI$ at time $t$ in SM.

       We present now a characteristic case to show how the preceding observations are
       used to define the various rates, $r$,  contributing to the equations
      (we omit the details of the whole construction of the system because the procedure
       to be applied to the other terms is similar).

      Let
     $r$ be the rate at which $\II$ grows at the expense of the decrease of $\SI$.
     For global contacts $r_{\EG}=\beta\I$.
     Let $ab$ be a pair of type $SI$ at a fixed time $t$ such that $U_a=I,\ U_b=S$.
    The rate of local infection for this pair is the product of $\beta$ and the fraction of infected individuals in the four nearest neighbors
    of $b$.  This fraction can be regarded as  $(1 + 3p_{1}  )/4$ , where $p_1$ is the probability that a fixed triplet $abc$
    ($c$ being a fixed nearest neighbor of $b$ different from $a$)
     is in situation $(U_a=I,U_b=S,U_c=I)$, given that $(U_a=I,U_b=S)$.
      We have

       $$p_1=\frac{P_{t}(U_a=I,U_b=S,U_c=I)}{P_{t}(U_a=I,U_b=S)} $$

       Then, using ~(\ref{lebo}) and $P_{t}(U_a=I,U_b=S)=P_{t}(U_a=S,U_b=I)$ 

    $$p_1\simeq \frac{P_{t}(U_a=S,U_b=I)}{P_{t}(U_a=S)}$$

         which corresponds to  $\SI/4\S$ in  DA. Then we take

         $$r_{\EL}= \beta\big{(}\frac{3\SI}{16\S}+\frac{1}{4}\big{)}$$
       and $(pr_{\EG}+qr_{\EL})\SI$ is incorporated as a substracting term 
       in the fifth equation and as an adding term
in the seventh equation of the system (6), that we present below.

      $$\frac{d\S}{dt}=-p\beta\S\I-\frac{q\beta\SI}{4}+\mu\I+\mu\R  \qquad\qquad\qquad\qquad \qquad\qquad\qquad\qquad $$
      $$\frac{d\I}{dt}=-\gamma\I-\mu\I+p\beta\S\I+\frac{q\beta\SI}{4} \qquad\qquad\qquad\qquad\qquad\qquad\qquad\qquad $$
     $$\frac{d\R}{dt}=-\mu\R+\gamma\I \qquad\qquad\qquad\qquad\qquad\qquad\qquad\qquad\qquad\qquad\qquad\qquad\quad$$
     $$ \frac{d\SS}{dt}=-2p\beta\I\SS-\frac{3q\beta\SI\SS}{8\S} +\mu\SI+\mu\SR \qquad\qquad\qquad\qquad\qquad\qquad\qquad\quad$$
\begin{eqnarray}
\frac{d\SI}{dt}=-p\beta\I\SI-q\beta\big{(}\frac{3\SII}{16\S}
               +\frac{\SI}{4}\big{)}-\gamma\SI-\mu\SI  +2p\beta\I\SS \qquad \qquad \\
            +\frac{3q\beta\SI\SS}{8\S}+2\mu\II+\mu\IR
            \qquad \qquad \qquad \qquad \quad \
            \qquad \qquad \qquad \qquad   \nonumber
\end{eqnarray}
      $$\frac{d\SR}{dt}=-p\beta\I\SR-\frac{3q\beta\SI\SR}{16\S}-\mu\SR+\gamma\SI+\mu\IR+2\mu\RR \qquad\qquad\quad   $$
      $$\frac{d\II}{dt}=-2\mu\II-2\gamma\II+p\beta\I\SI+q\beta\big{(}\frac{3\SII}{16\S}+\frac{\SI}{4}\big{)} \qquad\qquad\qquad\qquad  $$
     $$ \frac{d\IR}{dt}=-2\mu\IR-\gamma\IR+2\gamma\II+p\beta\I\SR+\frac{3q\beta\SI\SR}{16\S} \qquad\qquad\qquad\qquad\qquad $$
      $$\frac{d\RR}{dt}=-2\mu\RR+\gamma\IR \qquad\qquad\qquad\qquad\qquad\qquad\qquad\qquad\qquad\qquad\qquad\quad      $$

   For better visualization of the dynamics, we presented the initial
system of nine differential
equations. By using the equalities ~(\ref{L1}) to ~(\ref{L4}) we reduced the system
     to five ones for the numerical resolution.


\section{Results and Discussion}
\label{sectionRES}
In this section we compare the results obtained from 
the stochastic model (SM) simulations
with the results obtained using the deterministic approximation (DA).

We consider L=800 which mimics a city of $N$=640,000 inhabitants.
For pertussis we take 
 $\mu=1/(50$ years), $\gamma=1/(21$ days) and $\beta=0.8$ 1/day, 
which are standard parameters for SIR description 
of the disease in the pre-vaccine era \cite{rozhnova}. 
For  rubella we take $\mu=1/(50$ years), $\gamma=1/(18$ days) and 
$\beta=0.389$ 1/day \cite{keeling}.
We consider values of $p\geq0.2$ since in ref.\cite{dottori} 
it was observed that for lower values of $p$
the probability of establishment 
and survival of the steady state (representing the endemic state
of pertussis in prevaccine era) is very low. In the case of rubella,
this probability is very low even for $p=0.2$.
In this work we are interested in considering
values of the parameters that may represent real systems, however,
we have included in our study the case $p=0.2$ for rubella 
for comparison purposes.

In DA differential equations are integrated using Euler
algorithm with a time step 0.01 days that we have checked to
give accurate enough results for the problems studied.

\subsection{Epidemic behavior}
\label{subsecEB}
We first study the epidemic spread of the disease
for the case $p$=0.4.
We consider the evolution of the system when a single
infected individual is introduced in a population of $N$-1 susceptible
individuals.
From this initial condition, 5,000 independent runs were performed
for SM with different sets of random numbers.
For DA we take 
{\bf X$_0$}=$(1-1/N, 1/N, 0, 2-4/N, 4/N, 0, 0, 0, 0)$
as initial condition.

In Fig.1a we compare the results for three representative samples of SM 
simulations and
 DA for pertussis. In Fig.1b the same comparison is shown
for rubella. In both cases DA correctly predicts the qualitative
spread of the epidemic and also gives a  quantitatively good
prediction for the maximum fraction of infected individuals
reached during the course of the epidemic. 
In Fig.1c and 1d we compare the result for DA with the average
of the 5,000 runs of SM simulations. The discrepancy in the maximum of the fraction of
infected individuals in this case is caused because in the average
for the SM runs we included the extinctions. 
At the very beginning of the epidemic spread there is one infected
individual and the total rate of infection is $\beta$, while
the rate of recovery or death of the infected individual
is $\gamma \,'=\gamma+\mu$. As $\beta > \gamma\,'$ for both
diseases, DA always predicts an initial increase in the fraction
of infected individuals. In the SM
simulations there is a positive probability that recovery occurs
before infection, 
and since there is only one infected individual
at the beginning, recovery implies extinction.
The fraction of extinctions is greater in the case of rubella
as $\gamma\,' / \beta$ is greater as well.
At the first steps of the dynamics, the value of $p$ is irrelevant
as the infected individual can only contact susceptible ones
regardless of whether the contacts are local or global. 
This can be seen
in Fig.1e and 1f where SM simulations and DA predict an exponential
increase in the fraction of infected individuals at a rate
$\beta - \gamma -\mu$. An interesting result predicted by SM simulations
(and very well aproximated by DA) is the approximately exponential
growth of the epidemic a few days after its beginning
with an                          
exponent that differs from the initial one.
This can be concluded from the linear behavior
observed in the curve of Fig.1e from 5 to 15 days, and
in Fig.1f from 10 to 30 days. 
Exponential fits to the curves
in the mentioned ranges give exponents
0.61 and 0.63 for SM and DA for pertussis, and 
0.267 and 0.272 for SM and DA for rubella. 
\begin{figure}[ht]
\subfigure[ ] {
\includegraphics[width=6.8cm,height=4.8cm]{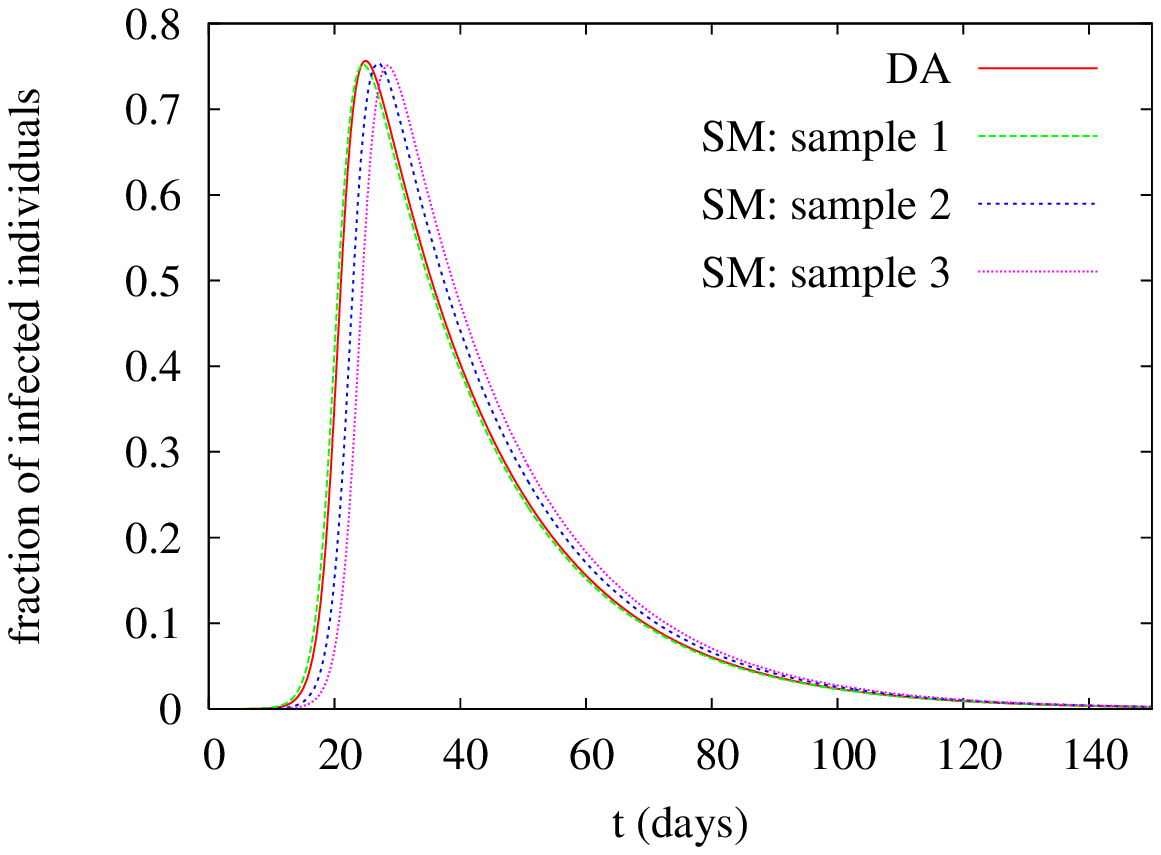}}
\subfigure[ ] {
\includegraphics[width=6.8cm,height=4.8cm]{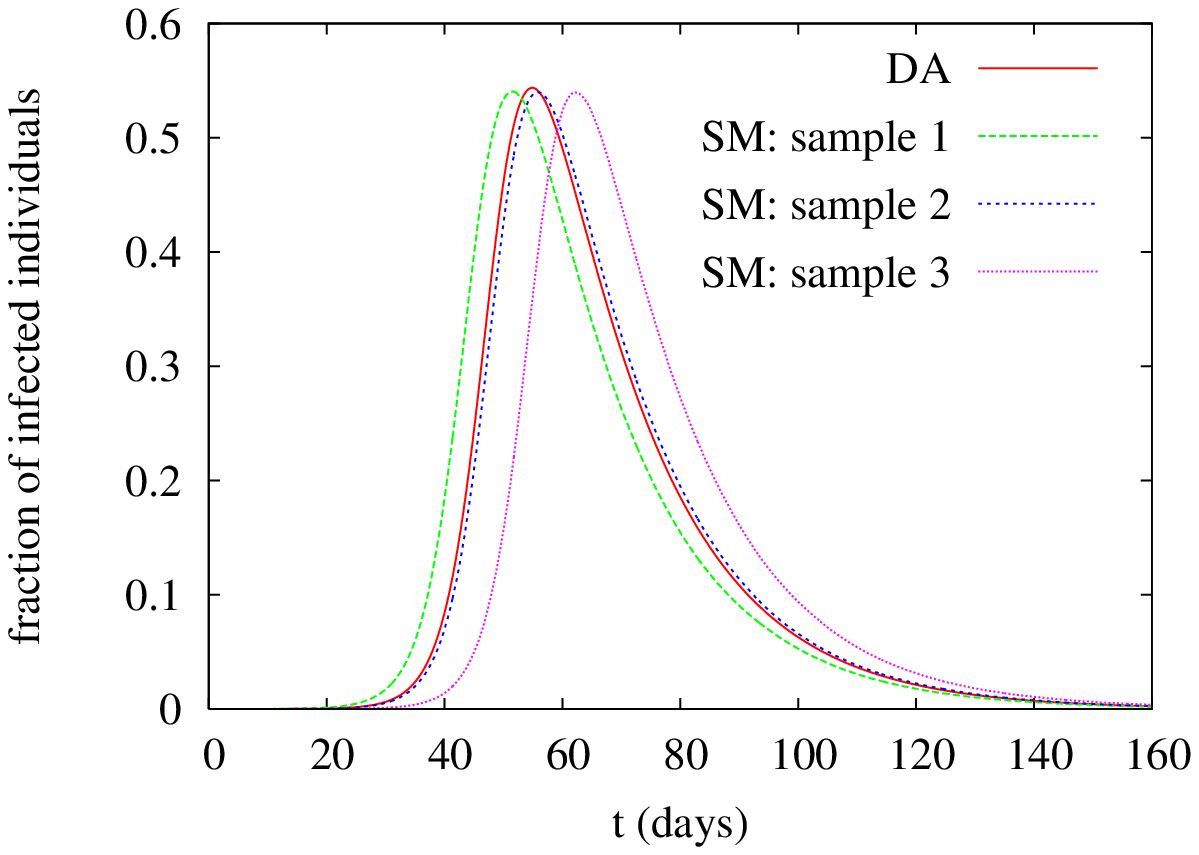}}
\subfigure[ ] {
\includegraphics[width=6.8cm,height=4.8cm]{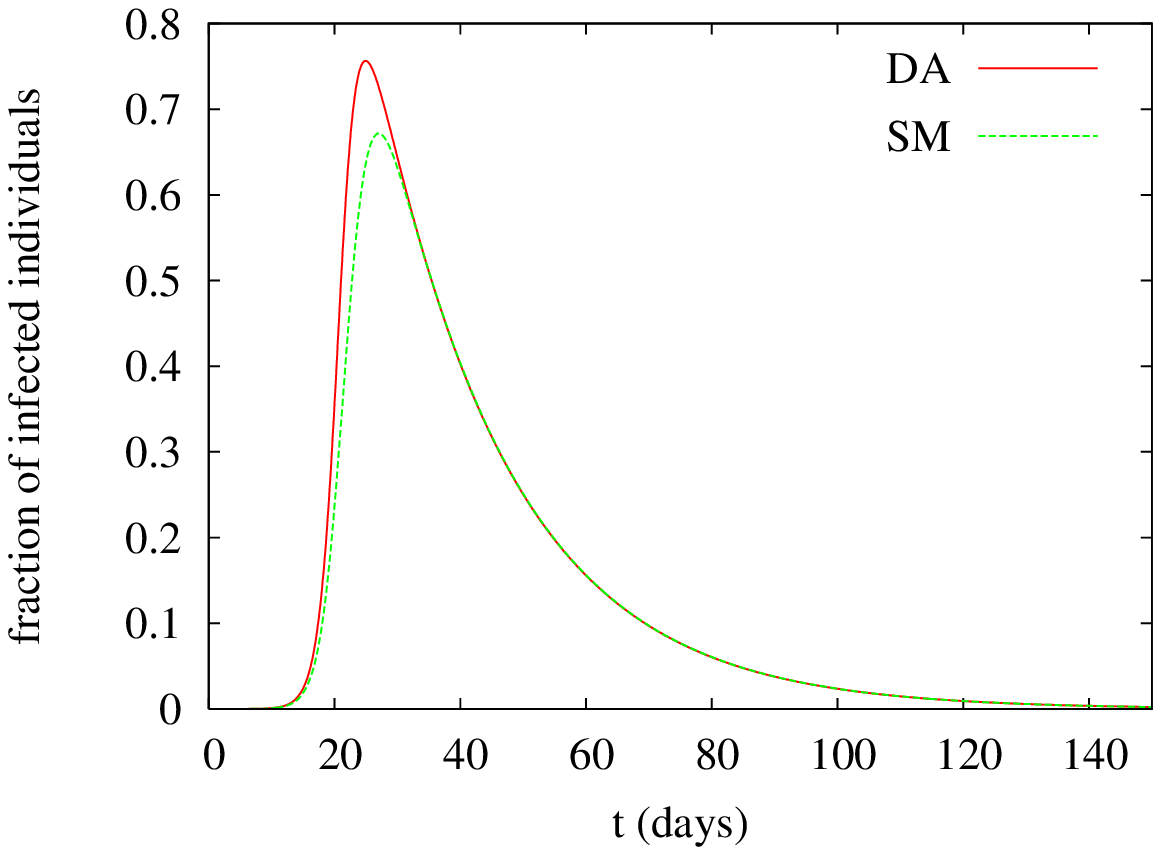}}
\subfigure[ ] {
\includegraphics[width=6.8cm,height=4.8cm]{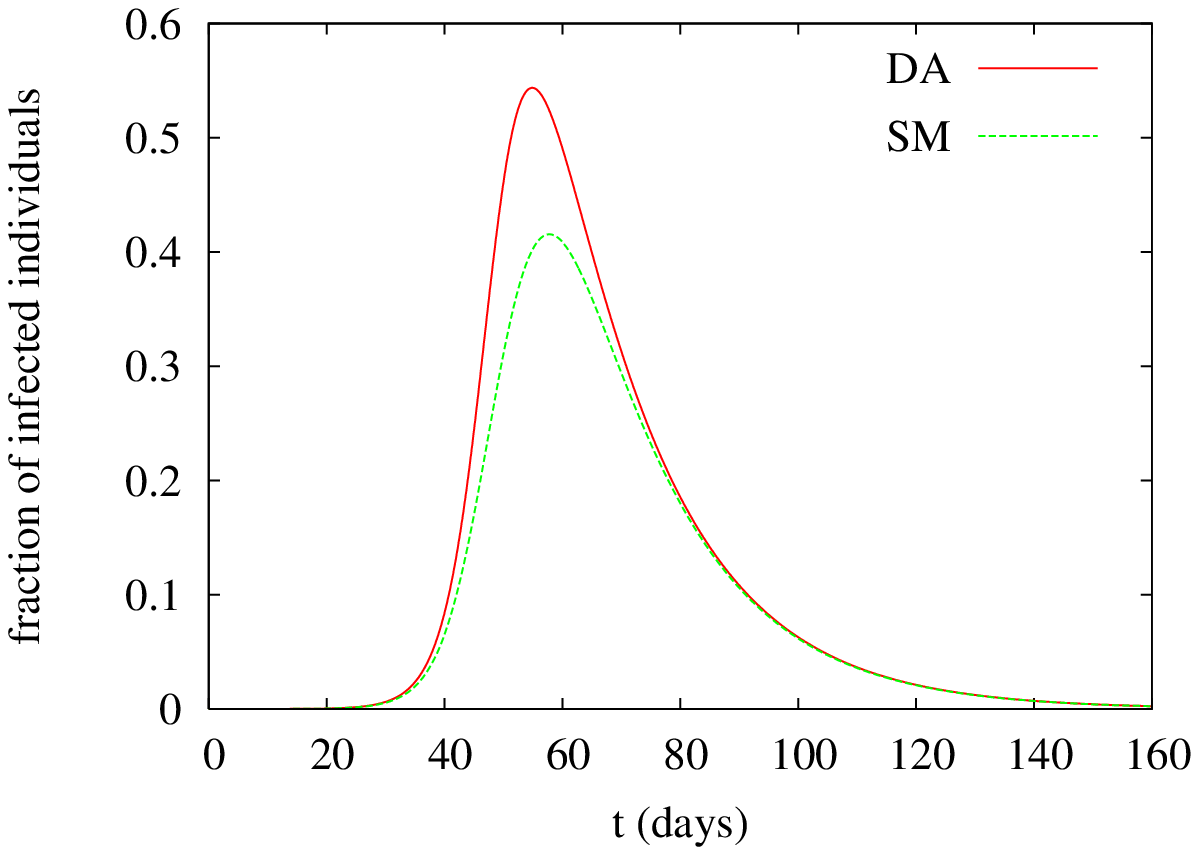}}
\subfigure[ ] {
\includegraphics[width=6.8cm,height=4.8cm]{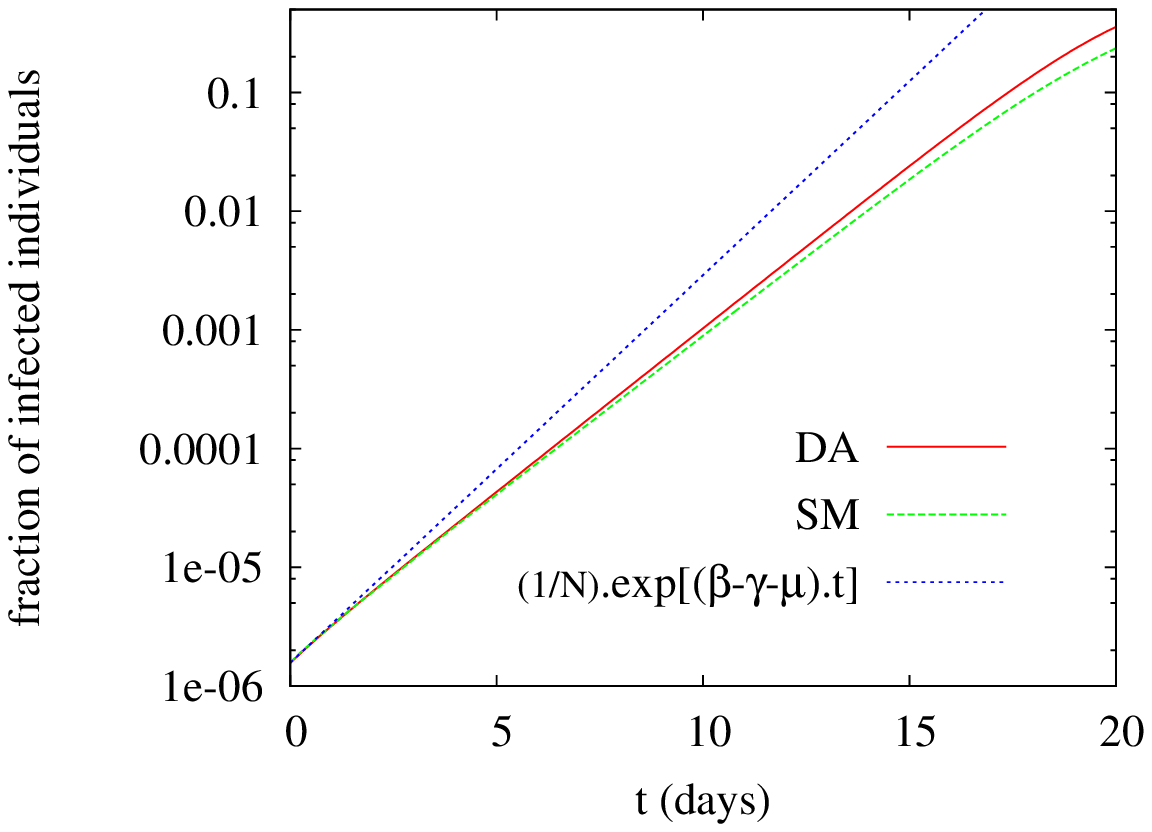}}
\subfigure[ ] {
\includegraphics[width=6.8cm,height=4.8cm]{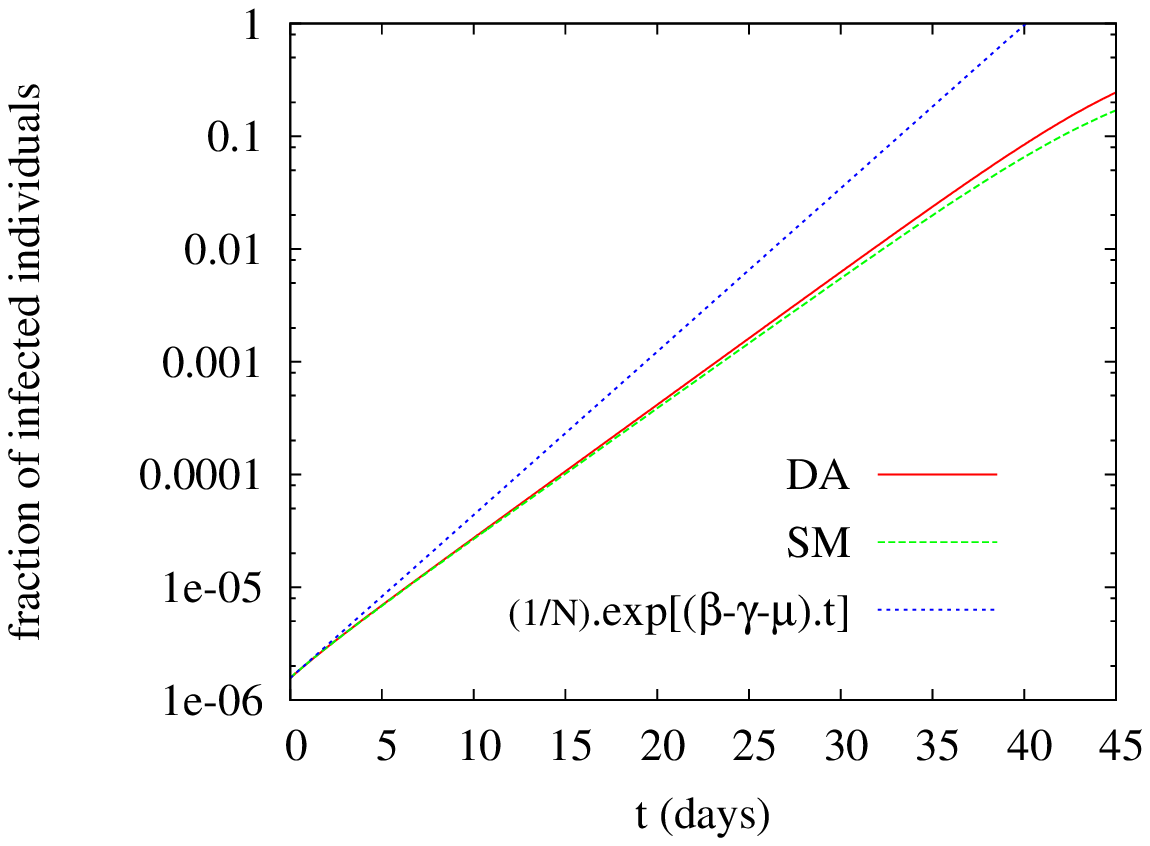}}
\caption{Fraction of infected individuals 
as a function of time (t) during the epidemic 
spread for $p=0.4$. 
(a) Comparison of DA result with SM simulations for pertussis. 
For SM three different
samples obtained with the same initial condition and different
stochastic evolutions are shown; 
(b) the same as (a) for rubella;
(c) comparison of DA result with the mean value function of  
    SM averaged over 5,000 samples for pertussis; 
(d) the same as (c) for rubella;
(e) initial growth of the epidemic for pertussis (detail of figure (c));
(f) initial growth of the epidemic for rubella (detail of figure (d)).
}
\end{figure}

In order to characterize this behavior as a function of $p$,
we notice that in this exponential regime the time dependence
of the fraction of infected individuals
may be written as $c.\exp[(\beta_{exp}-\gamma -\mu) t]$. That is to say,
the same as in the classical SIR model (with only
global contacts) but with an effective contact rate, $\beta_{exp}$, that
is different from $\beta$ because of the presence of local contacts.

It is possible to obtain an analytical approximate expression
to estimate  $\beta_{exp}$ 
using DA equations as follows.
At the beginning of epidemic growth, when $\I \ll 1$, we 
have that $\R$ , $\SI$ and $\SR$ are also $\ll 1$.
So, we drop all the second order terms in 
$\I, \R, \SI$ and $\SR$,
and also the terms $\II, \RR, \IR$.
As $\S=1-\I-\R$, we may write the 2nd equation of system (6):
\begin{equation}
\label{didtap}
\frac{d\I}{dt}\simeq -\gamma\I-\mu\I+p\beta\I+\frac{q\beta\SI}{4} 
\end{equation}

If $\I(t)$ presents an exponential behavior, 
$d\I / dt$ should be proportional to $\I$, so, we assume
that in this exponential regime: $\SI \simeq K \I$, $K$
being a constant.
Expression (\ref{didtap}) may now be written as:
\begin{equation}
\label{didtap2}
\frac{d\I}{dt}\simeq -\gamma\I-\mu\I+ \left( p+\frac{q K }{4} \right) \beta \I 
\end{equation}
where $(p+ qK/4) \beta$ plays the same rol as $\beta$ in the
exponential growth when only global contacts are present.
So, we define $\beta^{DA}_{exp}=(p+ qK/4) \beta$.

Taking $\SI \simeq K \I$ and replacing it in
the 2nd and 5th equations of system (6), 
we may obtain the value of $K$ and compute:
\begin{equation}
\label{betaexp}
\beta^{DA}_{exp} \simeq \frac{1}{4} 
                   \left( 1 + p + \sqrt{1+10 p -7 p^2} \right) \beta
\end{equation}

An interesting point here is that, at this level of approximation,
DA predicts that $\beta^{DA}_{exp} / \beta$ only depends on $p$ and
not on the other parameters characterizing the disease
transmission.
In Fig.2 we plot this relation for DA (from eq.(\ref{betaexp}))
and for SM (obtained from exponential fits for different $p$-values) for
pertussis and rubella. The agreement between DA and SM is good,
but it becomes worse for lower values of $p$. 
However, even for $p=0.2$ (when the agreement of SM and DA 
values for $\beta_{exp}$ is not so good) 
$\beta^{SM}_{exp} / \beta$  
accurately verifies the DA prediction
of being dependent only on $p$ for pertussis and rubella.

\begin{figure}[ht]
\includegraphics[width=6.8cm,height=5.0cm]{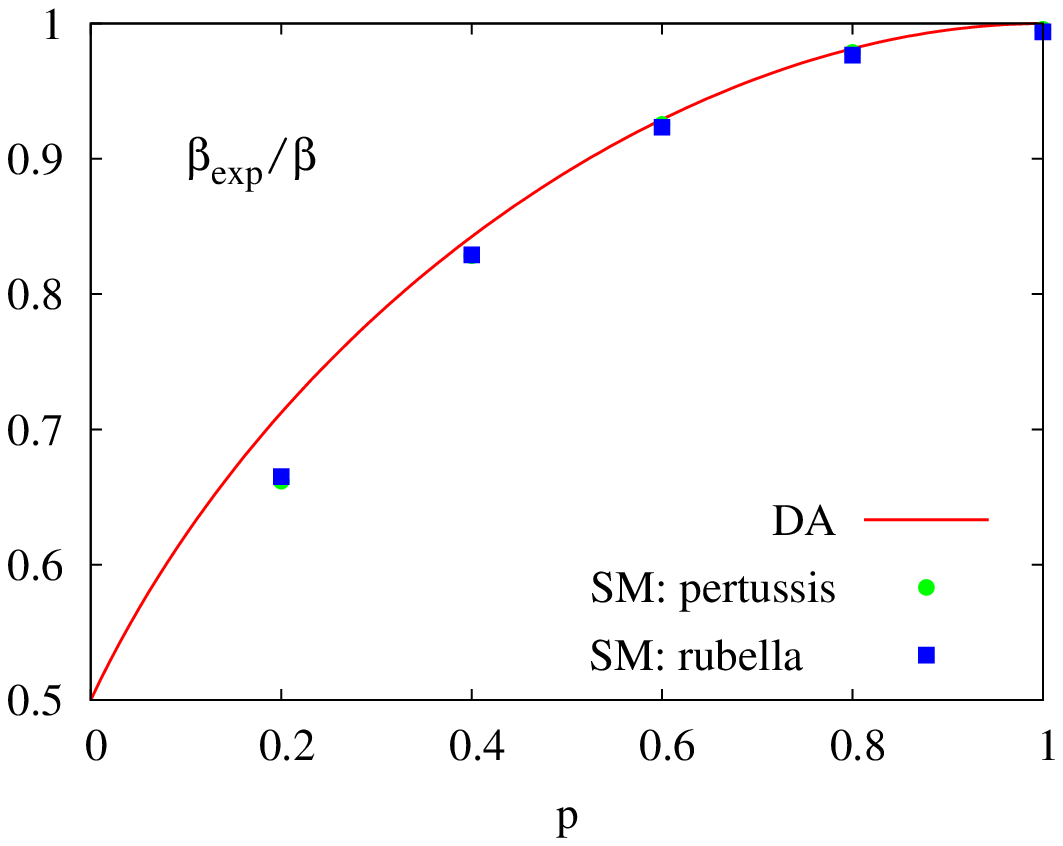}
\caption{\label{figbexp} 
Comparison of SM and DA predictions for the exponential
growth regime in an epidemic as a function of $p$.
The computation
of $\beta^{DA}_{exp} / \beta$ is performed through expression
\ref{betaexp}\ for DA. For SM, $\beta^{SM}_{exp}$ is obtained
by fitting the exponential regime of the curve
obtained for the average fraction of infected
individuals as a function of time 
to $ c \exp[(\beta^{SM}_{exp} -\gamma -\mu)t]$ for each 
case, where $c$ is a constant.}
\end{figure}

\subsection{Stationary behavior}
\label{subsecSB}
For the SM presented in Section \ref{sectionSM}\ and a finite value of
$N$, the only fixed point
(stationary equilibrium) corresponds to the case where all the individuals
on the lattice are susceptible. For the stochastic nature of the system,
sooner or later a fluctuation leads the number of infected individuals
to zero
and there is no process that produces new infected people 
if there are none. But for large values of $N$ (as the one taken in the present
study) the system may fluctuate
for a long time around a quasi-stationary state (QSS) before extinction.
The definition and properties of such a state have been addressed
in other contexts from a mathematical point of view
\cite{daroch-seneta,nasell}
or with empirical approaches \cite{dickman,zheng}.
In the present work we adopt the empirical strategy developed 
in ref.\cite{dottori} where the QSS of the same SM
defined in Section \ref{sectionSM} is studied.
We define a time window 
($t_a,t_b$) and generate (for each system studied) 
more than 20,000 independent samples 
keeping those that do not become extinct before $t_b$.
We compute the averages of susceptible and infected individuals
over the surviving samples and observe that in the ($t_a,t_b$) interval
 they remain constant in time
within a given precision.
We take $t_a$=20,000 days and $t_b$=40000 days; with these
values the considered averages are independent of
the initial conditions (as $t_a$ is large enough so
that correlations in the system may be established) and 
enough statistics is obtained to compute averages
with a relative precision of 0.01.
(For details of the definition of the QSS
and characterization of the fluctuations 
for the pertussis case see ref.\cite{dottori}.)

In the case of DA, stationary values are defined 
as the values of the variables that make all the derivatives
in equations (6) zero. They were obtained 
numerically solving the equations 
from a given initial condition, {\bf X$_0$}, until 
{\bf X}$(t)$ is constant in time with a precision 0.00001.
For {\bf X$_0$}
we take for $\S, \I$ and $\R$ the values corresponding 
to the stationary solution of the SIR model and 
complete the remainding values
of {\bf X$_0$} as in the uncorrelated case. For example,
$\SI$ is initialized as $4 \S \I$ because it is twice
the probability that a given pair is of type
$SI$ whose value is $2 \S \I$ in the uncorrelated case.

\begin{figure}[ht]
\subfigure[ ] {
\includegraphics[width=7.8cm,height=5.5cm]{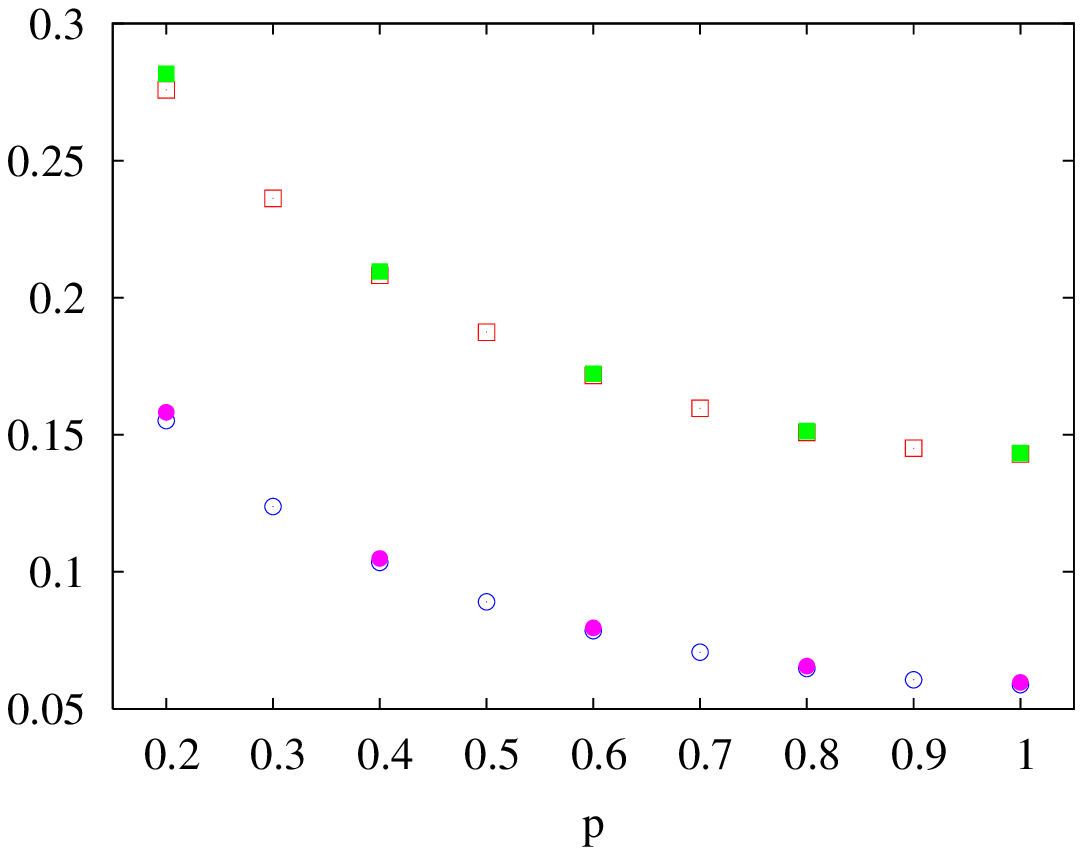}}
\subfigure[ ] {
\includegraphics[width=7.8cm,height=5.5cm]{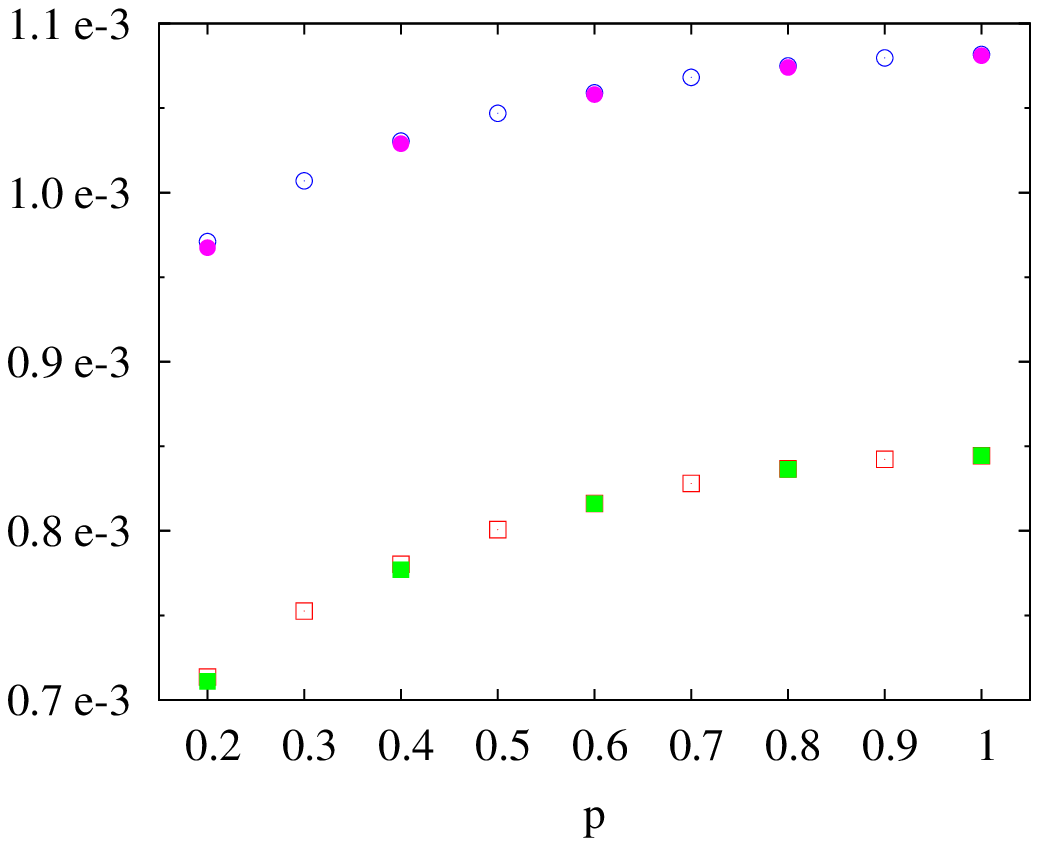}}
\subfigure[ ] {
\includegraphics[width=7.8cm,height=5.5cm]{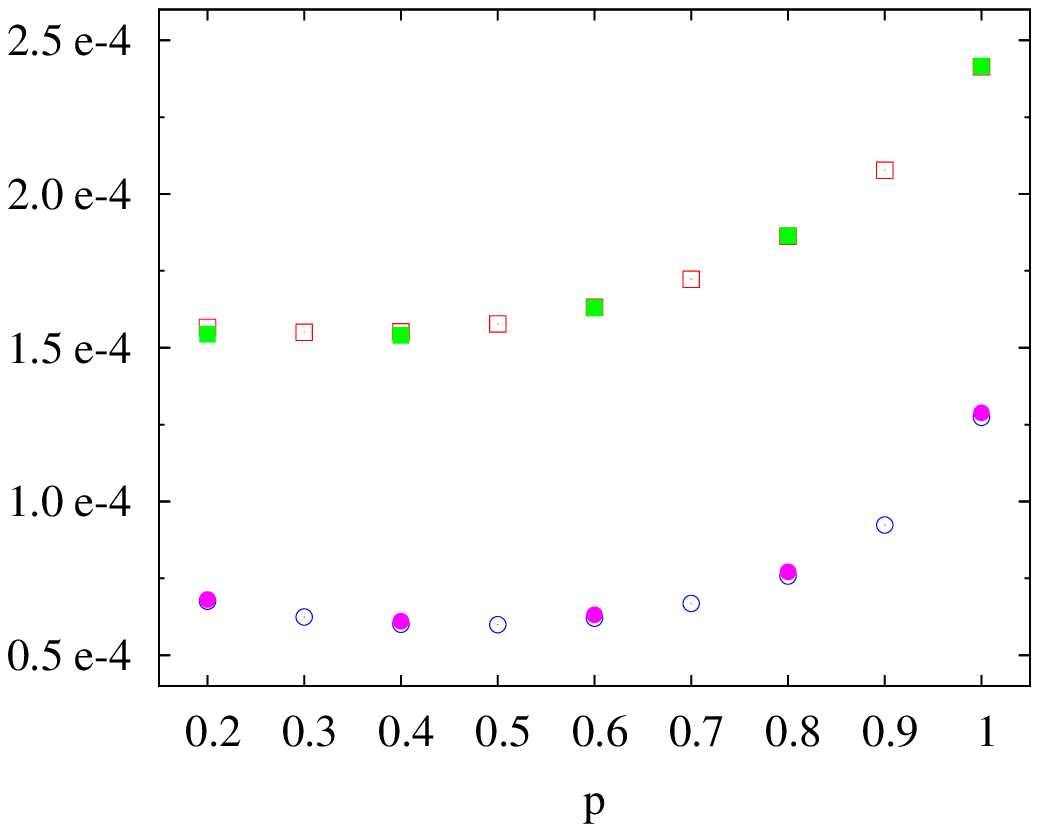}}
\caption{\label{figstat} Comparison of DA and SM results for
the stationary behavior. (a) Fraction of susceptible individuals; 
(b) fraction of infected individuals, and (c) fraction of $SI$ pairs on the
lattice. 
Circles correspond to pertussis and squares to rubella. 
Empty symbols correspond to DA calculations and 
filled symbols to the estimated average magnitudes in SM.
}
\end{figure}

In Fig.\ref{figstat}a and b we compare the DA stationary values
obtained for the fractions of susceptible and infected individuals for 
pertussis and rubella with the values of these observables
averaged over the samples in the QSS of SM. 
Results for both diseases show an excellent agreement between SM and DA.
In Fig.\ref{figstat}c we make the same comparison for the fraction 
of $SI$ pairs on the lattice. 
The agreement is remarkable considering
that this is a correlated quantity directly involved in the DA
to SM. 

\subsection{System response to a change in $p$}
\label{subsec_p}
We here study the ability of DA to describe the system
response to a sudden change in the globality of the contacts
represented in the model by the parameter $p$.
We simulate disease transmission at the endemic state in a city
where, at a given time,
infective contacts among people change and become more local.
Such a change could occur because of several circumstances
that affect social activities.
For example, in Argentina, during the 2009 pandemic flu,
the authorities took preventive measures that 
included suspension of classes at school in July. The population was advised 
to stay at home whenever possible and to avoid crowded places. 
These measures aimed at reducing the possibility
of an epidemic outbreak during the winter.
Our purpose here is to simulate the consequence
that this  sudden reduction in globality of the contacts
may have on the disease transmission of other
infectious diseases that are at the endemic state.

We perform simulations with  SM for parameters
corresponding to pertussis in the prevaccine era and
$p=0.5$.
Once the QSS is stablished, at a given time (t=20,000 days)
we change the value of $p$ to 0.4, which
corresponds to a 20\% reduction in the globality of contacts.
In Fig.4a we compare the dynamic evolution
of the system for three samples obtained from SM simulations
and DA result.
To obtain the DA curve we take 
the {\bf X} corresponding to the stationary state
for $p$=0.5 as initial condition and at $t=20,000$ we change the value of 
$p$ to 0.4.

We observe that DA captures the fall down in the fraction
of infected individuals after lowering $p$ and the 
pronounced peaks observed in the following years.
The perturbation at $t=20,000$ days puts in phase
the oscillations for the different samples
that become out of phase again from the third peak.
DA also accurately reproduces the time elapsed between peaks.
Oscilations are exponentially damped out in DA
until the new stationary value is reached. 
In the case of SM, fluctuations may be very different from sample
to sample \cite{dottori}, but as we have seen in Section 
\ref{subsecSB}, once the QSS is reached,
the fraction of infected individuals oscilates
around a value very similar to that predicted by DA.

We repeated this study for the case that
the system is initially
at the QSS corresponding to $p=0.6$ and at $t=20,000$
$p$ is changed to 0.4. In this case, although  the final
state is the same, the change in the globality of contacts
is 33\%. In Fig.4b it can be seen that after the perturbation,
the system reaches a deep minimum followed by pronounced maxima
much higher than in the previous case when the change in $p$
was 20\%.
Again, DA gives a good description of the observed effects.
In this case, most of the samples obtained from the SM simulations
become extinct when the system is at the first and pronounced minimum.
This is not surprising and could be inferred 
from the DA result that predicts the number
of infected individuals to be below 15 in the 
time interval (20,400, 20,800 days)
for the population considered here.
Extinctions in the SM simulations are thus expected, given the 
fluctuations in the number of infected individuals 
observed in the dynamic behavior of the system.
\begin{figure}[ht]
\subfigure[ ] {
\includegraphics[width=7.8cm,height=5.5cm]{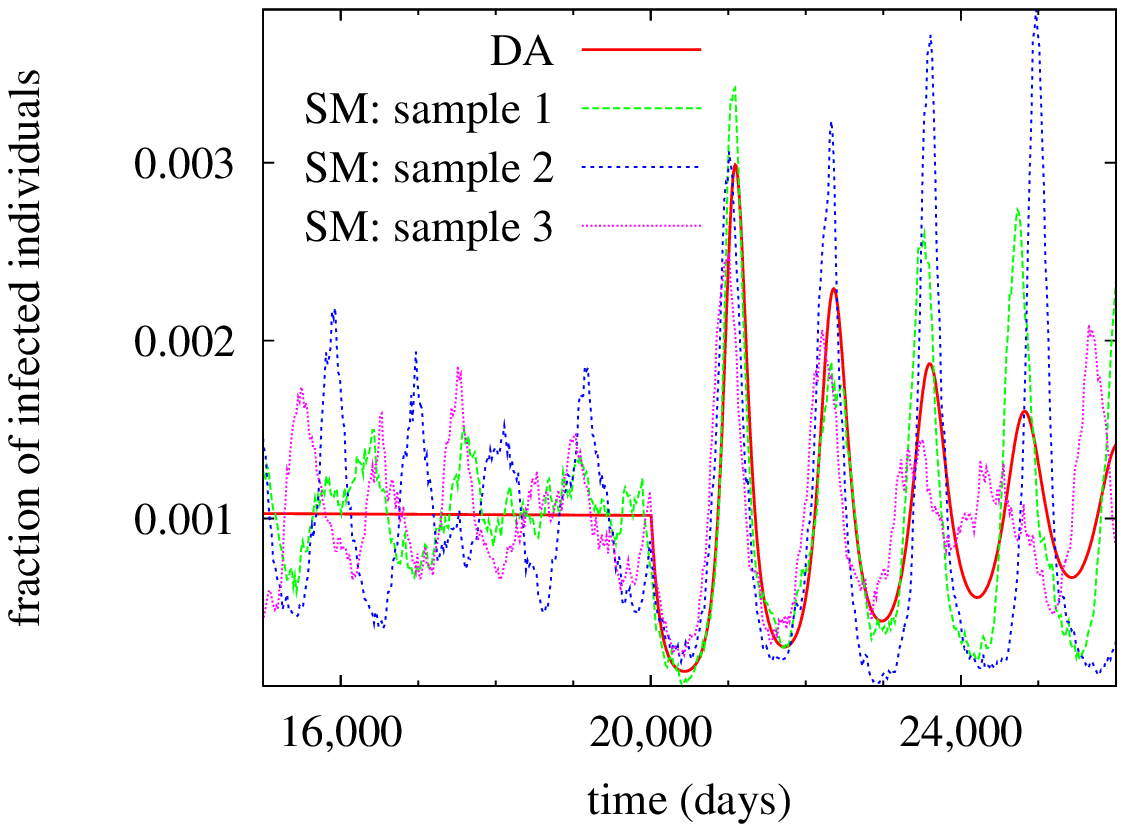}}
\subfigure[ ] {
\includegraphics[width=7.8cm,height=5.5cm]{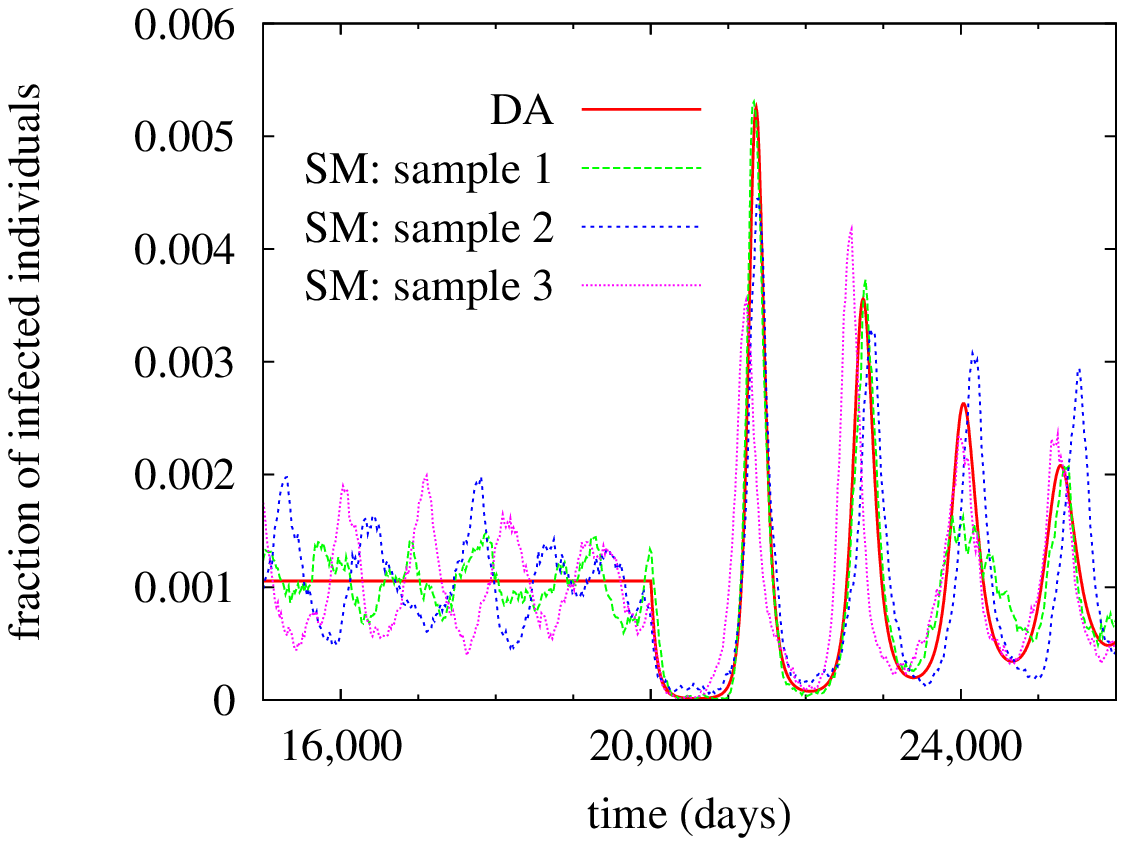}}
\caption{\label{fig_p} 
Time evolution of the fraction of infected individuals 
for pertussis when a sudden reduction in global contacts occurs.
(a) Comparison of DA result with SM simulations 
when the $p$ parameter is changed from 0.5 to 0.4 at $t=20,000$ days.
Initially the system is at the QSS for SM and at stationarity
for DA.
(b) The same as figure (a) but $p$ is changed from 0.6 to 0.4.
In this case, the three samples shown for SM are representative of the ones
that survive in the time window shown in the figure. The survival
samples in this case represent a fraction of 20\% from all the
generated samples.
}
\end{figure}

It is worth mentioning that
a dynamic behavior similar to that observed in Fig.4
has been found in a recent study of pertussis transmission
performed with an age-structured
deterministic model with nine epidemiological classes \cite{nos-epidemics}.
In that work, the authors proposed that a sudden change
in contact rates among individuals could cause
a dynamic effect with
the presence of deep valleys followed by sharp maxima
such as those recently observed in some US states.
Here we obtained a similar behavior for the disease transmission
as a consequence of having reduced  the degree of globality of the contacts
in a much more simple epidemiological model.

\section{Conclusions}

In this work we propose a deterministic approach (DA) for 
an SIR-type epidemiological model that includes global and local contacts
on a square lattice. When comparing our DA with simulations performed
with the stochastic version of the model (SM), we obtained
a very good agreement for qualitatively different scenarios 
of disease transmission and parameters
corresponding to pertussis and rubella in the prevaccine era.

It would be very valuable to increase the complexity of
the model used in the present work in order to 
study pertussis transmission in the vaccine era.
The introduction of pertussis vaccination in the model
forces one  to include
new epidemiological classes that account for individuals
that have partially (or totally) lost their 
immunity \cite{Hethcote9799, nos-adol-boost}
as it is well known that
immunity to pertussis is not lifelong \cite{wendelboe}.
We expect the good description obtained with DA for SM
to hold if a model with additional epidemiological classes 
is considered, since the DA to SM involves bassically
the description of local contacts that has shown to work
very well in different dynamical situations.         
Stochastic simulations for  a more
complex model could be much more time-consuming 
and DA might be the only possible approach to the problem.

\end{document}